# Self-Propulsion of Liquid Marbles: Leidenfrost-Like Levitation Driven by the Marangoni Flow

Edward Bormashenko[*a,b], Yelena Bormashenko[b], Roman Grynyov[a], Hadas Aharoni[b], Gene Whyman[a] and Bernard P. Binks[c]

[a]*Ariel University, Physics Department, P.O.B. 3, 40700, Ariel, Israel*

[b]*Ariel University, Chemical Engineering and Biotechnology Department, P.O.B. 3, 40700, Ariel, Israel*

[c]*Surfactant & Colloid Group, Department of Chemistry, University of Hull, Hull. HU67RX. UK*

[*]Corresponding author:

Edward Bormashenko

Ariel University, Physics Department, Chemical Engineering and Biotechnology Department

P.O.B. 3, Ariel 40700, Israel

Phone: +972-3-906-6134

Fax: +972-3-906-6621

E-mail: edward@ariel.ac.il

Contains ESI

**ABSTRACT**

Self-propulsion of liquid marbles filled with aqueous alcohol solutions and placed on a water surface is reported. The characteristic of velocity of the marbles is *ca.* 0.1 m/s. The phenomenon of self-propulsion is related to the Marangoni solutocapillary flow caused by the condensation of alcohol, evaporated from the liquid marble, on a water surface. The Marangoni flow in turn enhances the evaporation of alcohol from marbles. Addition of alcohol to the water supporting the marbles suppresses the self-propulsion. The propulsion of liquid marbles is mainly stopped by water drag. The velocity of the center of mass of marbles grows with the increase of the concentration of alcohol in a marble. The velocity of marbles' self-propulsion is independent on their volume.

**INTRODUCTION**

Autonomous locomotion (self-propelling) of droplets driven by various physico-chemical mechanisms attracted widespread attention of researches in the last decade.[1-9] Various mechanisms of self-propelling have been introduced including the use of gradient surfaces[1-3], involving the Leidenfrost effect[5-7], exploiting micro-structured surfaces[5-7]. Self-propelling of micro-scaled objects[10] and macroscopic bodies such as a camphor boat[11-13] was investigated. Self-propelling supported by solid[1-3] and liquid[14-15] surfaces was reported. An interest to self-propelling systems arises from the numerous applications, including lab-on-chip systems, targeted drug delivery and microsurgery.[16-17]

This paper reports the self-propulsion of liquid marbles filled with aqueous alcohol solutions deposited on a water surface. Liquid marbles are non-stick droplets coated with colloidal particles. When attached to the surface of a liquid droplets, colloidal particles allowed manufacturing the so-called liquid marbles, which are non-stick droplets, presenting an alternative to the lotus-inspired superhydrophobicity.[18-20] Liquid marbles demonstrated a diversity of promising applications, including encapsulation, microfluidics, cultivation of microorganisms, gas-sensing, miniaturized synthesis and water storing.[21-34] Liquid marbles could be actuated by various stimuli including chemical actuation[21-22] and electrical and magnetic fields[24, 28,29]. Our paper relates the effect of self-propelling to the interplay of the Leidenfrost effect and the Marangoni solutocapillary flow. The reported results open a way for micromanipulation of small quantities of liquids supported by liquid surfaces.

**EXPERIMENTAL**

Liquid marbles containing aqueous ethanol (Absolute, Dehydrated, water (KF): max. 0.2% w/w, supplied by Bio-Lab ltd, Israel) solutions were manufactured with the extremely hydrophobic fumed fluorosilica powder used to prepare oil liquid marbles reported earlier.[35] The primary diameter of these particles is 20-30 nm and they originate from hydrophilic silica after reaction with tridecafluoro-1,1,1,2-tetrahydrooctyltrimethoxysilane. The residual silanol content on their surfaces is 50% and the fluorine content is 10.9%.

Aqueous ethanol droplets of volume between 10 and 50 μL were spread on a layer of fumed fluorosilica powder situated on a glass slide. The slide was vibrated slightly, giving rise to the formation of liquid marbles. The maximum volume concentration of alcohol in water allowing manufacturing of liquid marbles was established as 85 vol.%. A typical marble containing 70 vol.% ethanol in water is depicted in Figure 1. A larger concentration of alcohol promoted the Cassie-Wenzel wetting transition and marbles were not formed, as discussed in ref. 36. The marbles are not hermetically coated by powder; they evaporate as discussed in detail in ref. 23. The lifetime of marbles enabling self-propulsion is *ca*. 2 min.

Marbles were placed on the surfaces of bi-distilled water (resistivity 2 *MΩcm*$^{-1}$ as measured with LRC-meter Motech MT 4090), aqueous alcohol solutions, silicone (PDMS) oil ($M_n$= 5600 supplied by Aldrich) and glycerol (analytical grade, supplied by Frutarom) under the following procedure. Marbles were placed on the bottom of glass Petri dishes with the diameters of 5 and 25 cm, and a liquid was gently added to the dish. This procedure avoided supplying the initial velocity to highly mobile liquid marbles. The motion of liquid marbles was registered from above with a rapid camera, Casio EX-FH20. Use of Petri dishes with a diameter of 25 cm enabled excluding of boundary effects, due to the meniscus formed in the vicinity of the dish skirting.

In some experiments, the marbles contained a 0.1 wt.% solution of phenolphthalein and NaOH was added to the water supporting the liquid marbles at a concentration of 0.1 wt.%. This was to detect if any of the marble contents entered the supporting water phase.

**RESULTS AND DISCUSSION**

Liquid marbles containing aqueous ethanol solutions were placed on the surface of various liquids contained in a Petri dish, as described in the Experimental Section. The marbles did not survive on the surface of the silicon oil and immediately burst, as discussed in ref. 37; they survived and floated motionlessly on the surface of glycerol, as shown in Figure 2 but moved rapidly rectilinearly, when placed on a water surface, periodically bouncing the skirting of the Petri dish as depicted in Figure 3. The same type of motion, continuing *ca*. 10 s, displayed in Figure 3, was observed on Petri dishes with the diameters of 5 and 25 cm.

The surface tensions of water and glycerol $\gamma$ are close, whereas their viscosities $\eta$ differ markedly: $\gamma_w = 72 \frac{mJ}{m^2}; \gamma_{gl} = 63 \frac{mJ}{m^2}; \eta_w \cong 10^{-3} \frac{kg}{m \cdot s}; \eta_{gl} \cong 1.5 \frac{kg}{m \cdot s}$. Thus very different behavior of marbles placed on water and glycerol surfaces may be at least partially related to the effect of viscosities of supporting liquids, as it will be discussed below.

The velocity of the motion established at straight sections of the displacement with the rapid camera is illustrated in Figure 4. It is noted that the collisions of the marbles with the skirting of the Petri dish are elastic and the velocity of a marble remains constant after dozens of collisions (see the movies, supplied in the attached ESI).

It should be emphasized that aqueous alcohol solutions filling the marbles did not touch the supporting liquid surface. This was proved by placing of a marble containing phenolphthalein on water solutions of NaOH (see also ref. 38). No coloring of water below was observed under the self-propulsion of marbles meaning that moving marbles are isolated from the supporting liquid by a vapor layer, as occurs in the Leidenfrost effect.[39-40] The complicated geometry of the vapor layer, depicted schematically in Figure 5(a) was treated in detail in ref. 41. However, the typical value of the velocity of the center of mass of marbles was $v_{cm} \approx 0.1 - 0.15 m/s$ which is an order of magnitude lower than the velocity observed for Leidendfrost water droplets.[39-40] We relate this observation to the action of a water drag as will be discussed below in detail below.

What is the physical mechanism of the self-propulsion of liquid marbles? Remarkably self-propulsion occurs under breaking the spherical symmetry of marbles, i.e. the liquid marble, which evaporation is spherically symmetric, has no inherent reason for running. Consider the spontaneous increase in evaporation of alcohol from the marble in the direction of –$x$ (recall that alcohol evaporates from a marble much faster than water), depicted in Figure 5(a). This increase will not only drive the marble in the direction of +$x$ but also will give rise to a fascinating instability, driving marbles. Indeed, alcohol, condensed on a water surface (as shown in Figure 5(a)) will decrease its surface tension, resulting in the Marangoni solutocapillary flow[42-45] in the direction of –$x$. This flow in turn enhances the evaporation, withdrawing vapor from the layer, separating the marble from the supporting liquid (as shown with a green arrow in Figure 5(a)). The introduced instability displaces a marble in a direction of $+x$.

If the self-propulsion of marbles is related to the evaporation of alcohol resulting in the Marangoni flow, it is reasonable to suggest that the increase of the alcohol concentration in the marble will give rise to the larger velocities of their locomotion. This tendency was observed experimentally, as shown in Figure 6, representing the velocity of the center of mass of marbles as a function of the alcohol concentration in a marble.

If it supposed that a marble is driven by a solutocapillary Marangoni flow, the driving force may by qualitatively written as $F_{dr} \approx |\Delta\gamma| 2\pi a = \beta |\Delta\gamma| a$, where $|\Delta\gamma|$ is the modulus of the jump (lowering) of the surface tension due to the condensation of alcohol on the water surface, $a$ is the radius of the contact area (see Figure 5(b)), and $\beta$ is the dimensionless coefficient. Thus, the motion of the marble may be described by the following equation:

$$m\ddot{x}_{cm} = \beta |\Delta\gamma| a - F_{fr} \qquad (1)$$

Where $\ddot{x}_{cm}$ is the acceleration of the center of mass of a marble, $m$ is its mass, and $F_{fr}$ is the total friction force acting on a marble, built from three components: $F_{fr} = F_{air} + F_{vapor} + F_{water}$, where $F_{air}$ is the friction exerted on a marble by air, $F_{vapor}$ is the friction due to the viscous dissipation taking place in the vapor layer, separating the marble from the liquid below and $F_{water}$ is the viscous drag due to the water under the marble motion.[40] $F_{air}$ scales as $\rho_{air} v^2_{cm} R l_{ca}$, where $\rho_{air} = 1.2 kg \cdot m^{-3}$ is the density of air, $R \sim 1$mm is the radius of the marble, $v_{cm} \cong 0.1 m/s$ and $l_{ca} \cong 2.7 cm$ is the so called capillary length.[40, 45-46] For a realistic estimation of $F_{air}$ we have $F_{air} \approx 3 \cdot 10^{-8} N$.

The viscous dissipation occurring in the vapor layer scales as $F_{vapor} \cong \frac{\eta_v v_{cm}}{h} R^2$, where $\eta_v \approx 1.7 \cdot 10^{-5} \frac{kg}{m \cdot s}$ is the viscosity of the vapor layer and $h \approx 100 \mu m$ is its thickness.[40] Thus, the estimation of this component of friction is $F_{vapor} \cong 1.7 \cdot 10^{-8} N$. The viscous drug of water may be estimated roughly as $F_{water} = N f_{stokes}$, where $N$ is the number of solid beads of powder in contact with water, and $f_{stokes}$ is the Stokes drag acting on a single bead. $N$ may be estimated as $N \approx \frac{\pi a}{2b}$ ($b$ is the radius of a bead, see Figure 5(b)), and the Stokes drag scales as $f_{stokes} \approx \xi \eta_w b v_{cm}$, where $\eta_w \cong 10^{-3} \frac{kg}{m \cdot s}$ is the water viscosity and $\xi \approx 10$ is the dimensionless coefficient. Thus, the plausible estimation of the water viscous drag is $F_{water} \cong \xi a \eta_w v_{cm} \cong 10^{-6} N$. The comparison of the components of the friction leads to the expected conclusion that the friction is mainly governed by the viscous drag of the liquid. This explains why marbles are motionless when placed on glycerol; the high viscosity of glycerol prevents their motion. This situation is quite different from the motion of Leidenfrost droplets, decelerated mainly by $F_{air}$.[40]

If marbles are driven by the Marangoni solutocapillary flow[42-45] their movement will be suppressed if alcohol is added to water supporting the marbles. This hypothesis was validated by a series of experiments performed with different concentrations of alcohol $c_s$ added to water supporting the marbles. Indeed, addition of alcohol decreased the velocity of marbles dramatically, as shown in Figure 7.

Considering Eq. 1 and the aforementioned analysis of the friction force yields the following expression for the stationary velocity of the center mass of the marble:

$$\beta |\Delta \gamma| a - \xi a \eta_w v_{cm} = 0 \ . \tag{2}$$

Thus, the scaling law for the stationary velocity of the center mass of the marble is supplied by:

$$v_{cm} \cong \frac{|\Delta\gamma|}{\eta_w}, \quad (3)$$

and it turns out that the velocity of the center mass of the marble is independent on its volume. This prediction was checked experimentally for liquid marbles in which their volumes varied from 10 to 50 µl. Figure 8, presenting the dependence of the stationary velocity of marbles on their volume, demonstrates that their velocity is independent of volume, justifying the scaling law given by Exp. 3. It seems instructive to estimate the value of $|\Delta\gamma|$ giving rise to the velocities of liquid marbles, observed in our experiments. According to Exp.3 we have for the sake of a very rough estimation: $|\Delta\gamma| \approx v_{cm}\eta_w \cong 10^{-4}\frac{mJ}{m^2}$. It is seen that relatively small change in the surface tension of supporting liquids brings into existence the velocities of marbles as high as 0.1 m/s.

It is noteworthy, that the marbles did not roll (rotate) when moved. This was demonstrated with experiments performed with 20 µl Janus-marbles, built from hemispheres coated by fumed fluorosilica powder and carbon black, as depicted in Figure 9A. The Janus-marbles were prepared as described in detail in Ref. 24. They moved rectilinearly, when placed on a water surface, as discussed above. The carbon black coated hemisphere served as a tracer (see Figure 9B), indicating the absence of rotation of marbles under their self-propulsion.

**CONCLUSION**

We report the self-propulsion of liquid marbles, containing aqueous ethanol solutions placed on a water surface. Use of very low surface energy fluoro-particles enabled manufacturing of marbles containing concentrated aqueous ethanol solutions. Marbles moved with a velocity of 0.12-0.15 m/s and bounced elastically from the skirting of the Petri dish. The self-propelling occurs under breaking of the spherical symmetry of marbles. We relate the effect to the instability involving the spontaneous increase of evaporation of alcohol from a marble leading to the Marangoni solutocapillary flow. The Marangoni flow in turn enhances evaporation of alcohol from the marble, under withdrawing vapor from the layer, separating a marble from the liquid support. Introducing alcohol into the underlying water suppressed the self-propulsion phenomenon. The velocity of the center of mass of marbles grew with the increase of the concentration of alcohol in a marble.

The mechanism of motion of liquid marbles is analyzed. Marbles are stopped mainly by the water drag, which is much larger than the other components of friction including the air friction. The stationary velocity of marbles is independent of their volume.

**Notes**

The authors declare no competing financial interest.

**ACKNOWLEDGEMENTS**

Acknowledgement is made to the donors of the American Chemical Society Petroleum Research Fund for support of this research (Grant 52043-UR5). The authors indebted to Mrs. Al. Musin and Mrs. E. Shapiro for their kind help in preparing this paper.

**AUTHOR INFORMATION**

Corresponding Author

*E-mail: edward@ariel.ac.il.

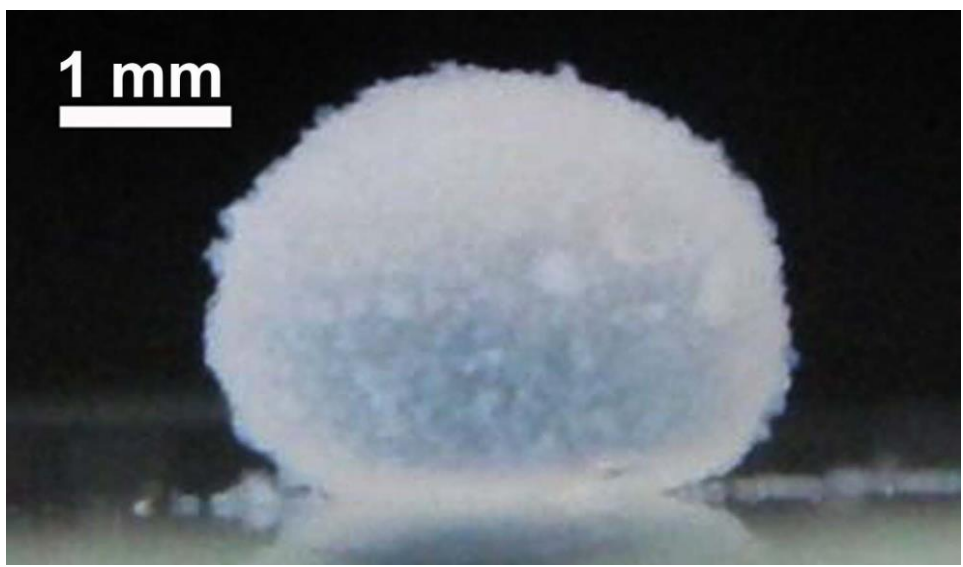

**Figure 1.** 10 μl liquid marble containing 70 vol.% aqueous ethanol solution, resting on a glass slide.

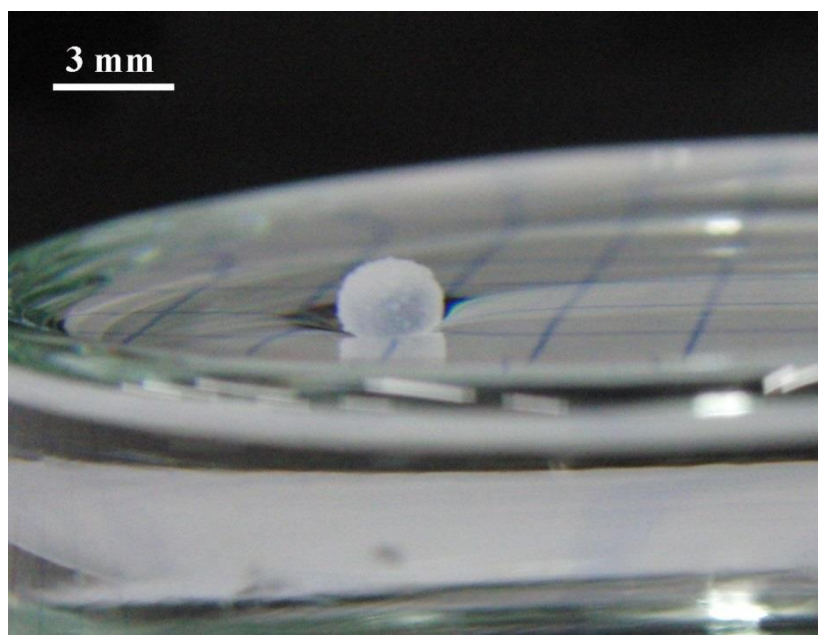

**Figure 2.** 10 μl liquid marble containing 70 vol.% aqueous ethanol solution floating on the surface of glycerol.

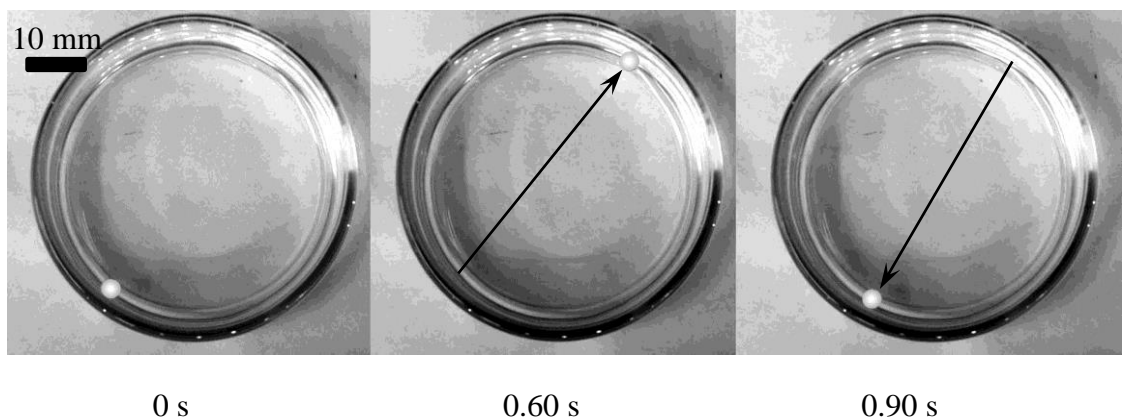

      0 s              0.60 s              0.90 s

**Figure 3**. Sequence of images demonstrating the motion of a10 µl liquid marble coated with fumed fluorosilica particles containing 70 vol.% aqueous ethanol solution placed on water.

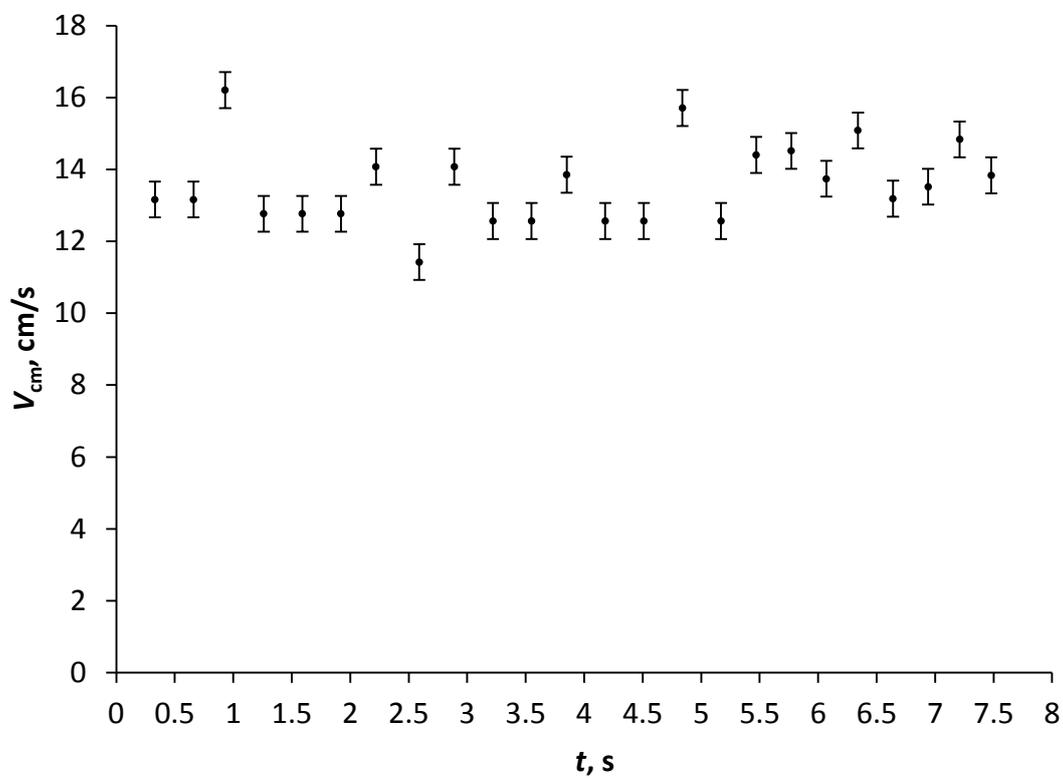

**Figure 4**. The velocity of the center of mass $v_{cm}$ of a 10 μl liquid marble coated with fumed fluorosilica particles containing 70 vol.% aqueous ethanol solution as a function of time. Every point corresponds to the bouncing of the marble with the skirting of the Petri dish.

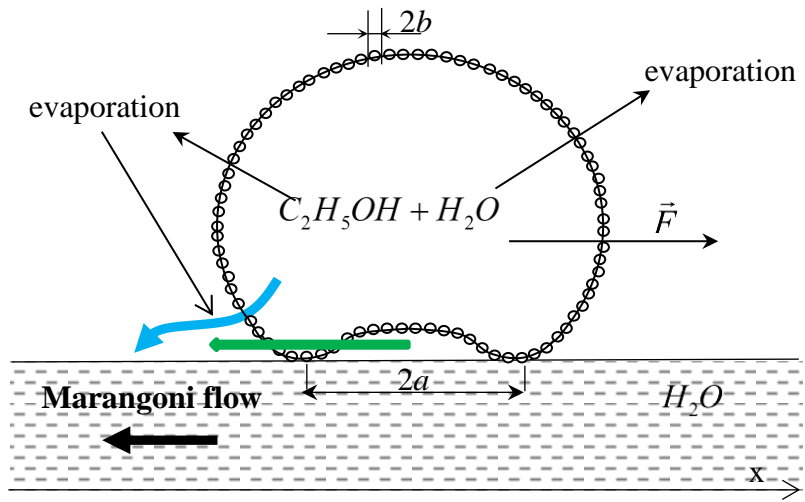

**Figure 5(a)**. Scheme illustrating the origin of the instability driving liquid marbles containing water/alcohol solution deposited on the water surface. The blue arrow shows spontaneous increase of the alcohol evaporation from a marble. The black arrow indicates the direction of the Marangoni flow, increasing in turn the evaporation of alcohol from the area beneath a marble (shown by the green arrow).

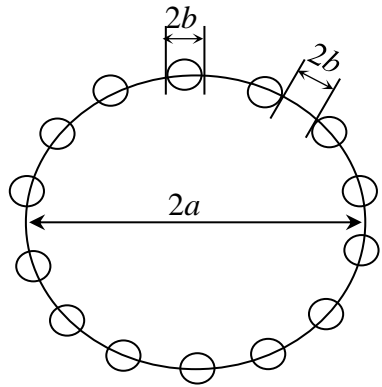

**Figure 5(b)**. Scheme of the contact area. Solid particles are separated by clearings of $2b$.

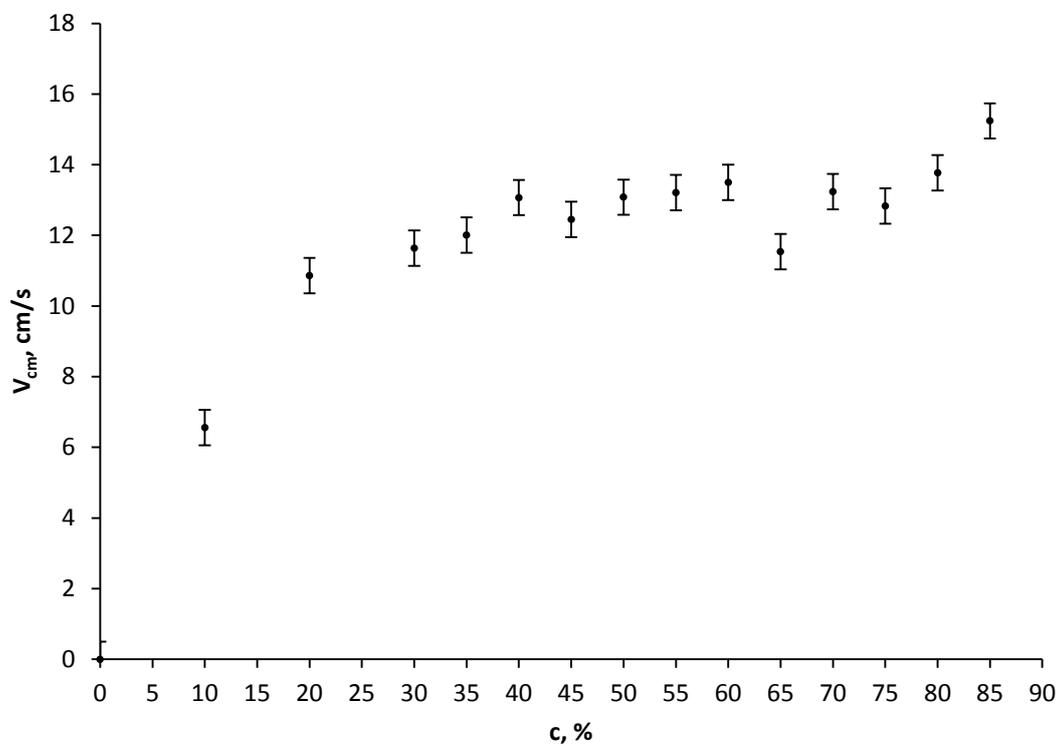

**Figure 6.** The velocity of the center of mass $v_{cm}$ of a 10 μl liquid marbles coated with fumed fluorosilica particles vs. the concentration of alcohol.

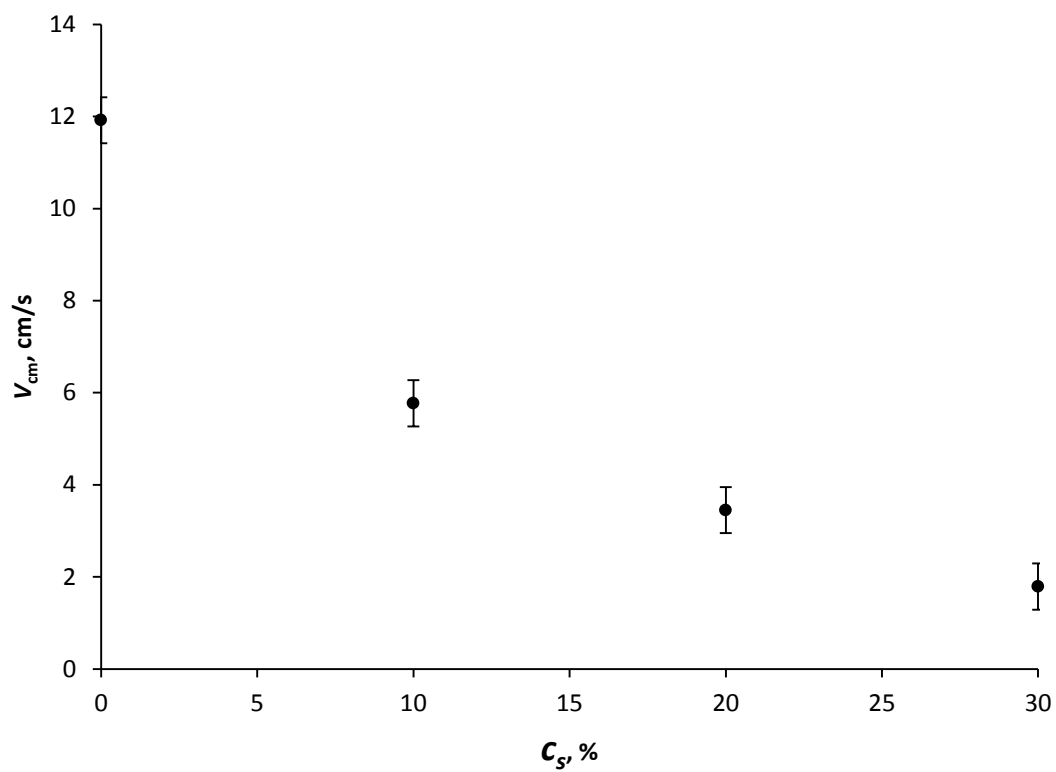

**Figure 7**. Initial velocity of 10 µl marbles containing 70 vol.% aqueous ethanol solution, placed on aqueous ethanol solutions surfaces *versus* the volume concentration of alcohol in the supporting solution $c_s$.

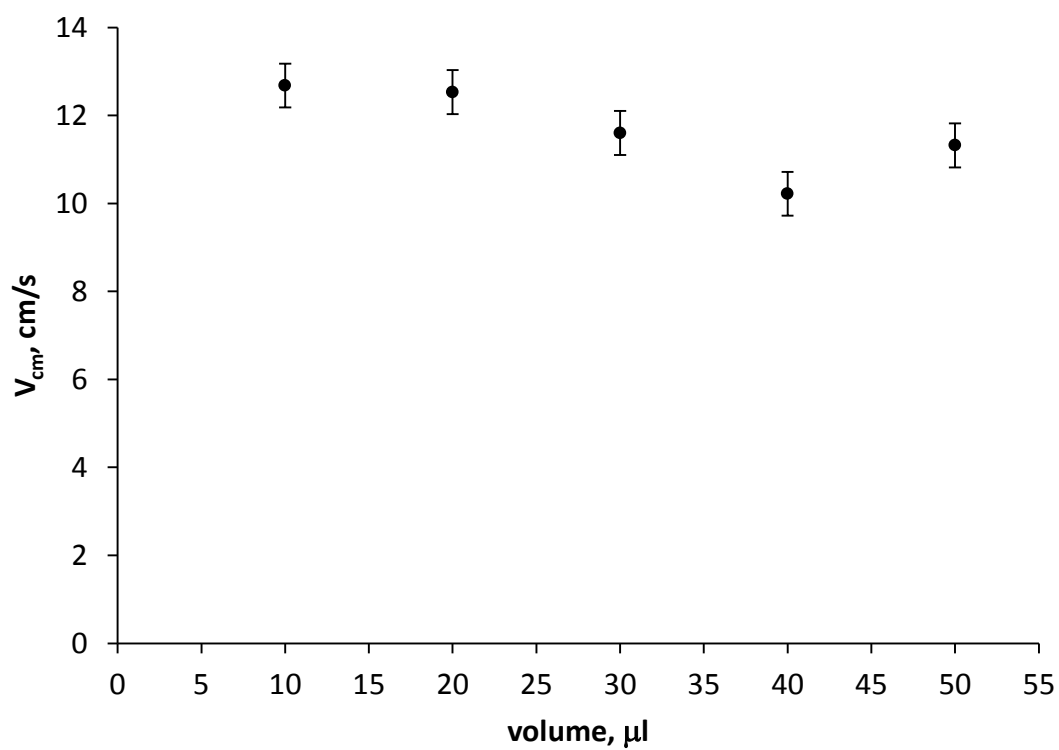

**Figure 8**. Stationary velocities of the marbles containing 70 vol.% aqueous ethanol solution established for marbles of various volumes.

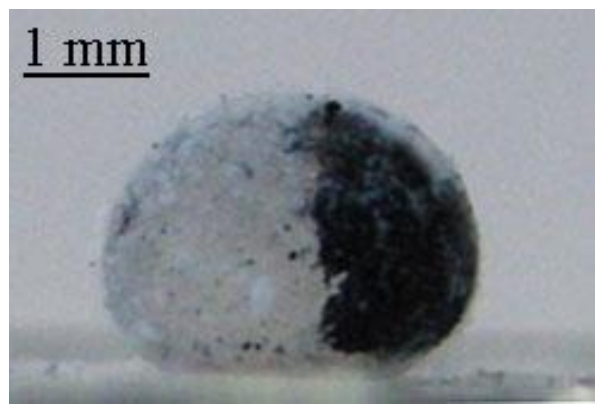

Figure 9A. 20 µl Janus marble coated with fumed fluorosilica powder and carbon black. The marble contains 37.5 vol.% aqueous ethanol solution.

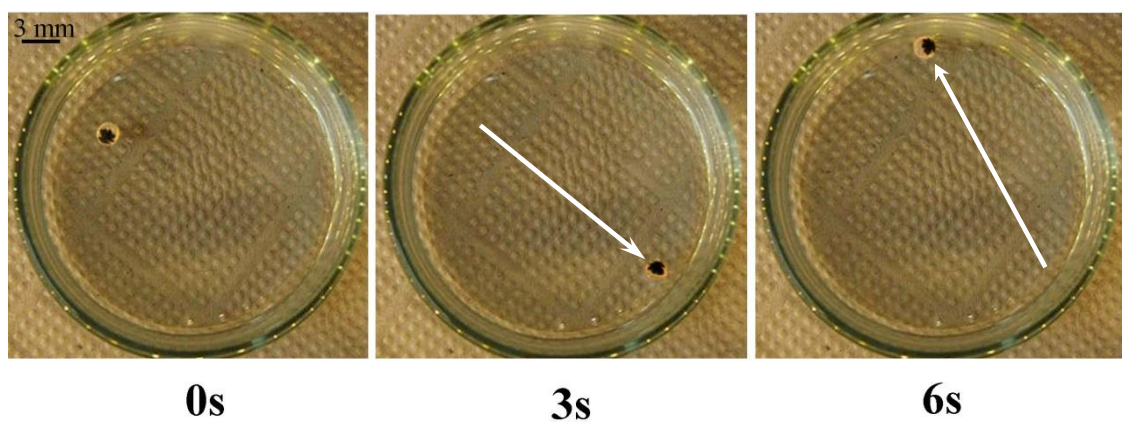

Figure 9B. Sequence of images demonstrating the self-propulsion of the 20 µl Janus marble containing 37.5 vol.% aqueous ethanol solution. The carbon black coated hemisphere works as a tracer.

TOC Image

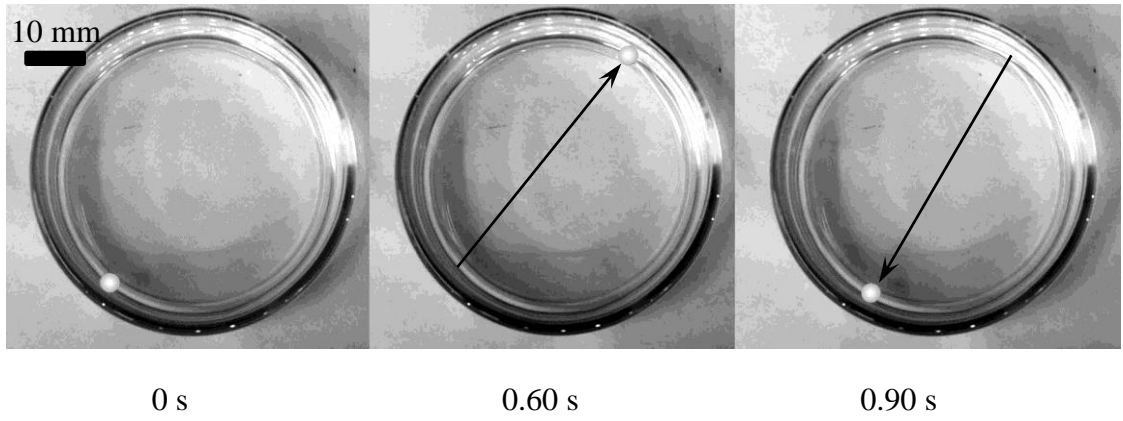

      0 s                 0.60 s               0.90 s